# Analysis of Learning-based Offshore Wind Power Prediction Models with Various Feature Combinations

Linhan Fang
Electrical & Computer Engineering
University of Houston
Houston, USA
lfang7@uh.edu

Fan Jiang
Electrical & Computer Engineering
University of Houston
Houston, USA
fjiang6@uh.edu

Ann Mary Toms
Electrical & Computer Engineering
University of Houston
Houston, USA
atoms2@cougarnet.uh.edu

Xingpeng Li
Electrical & Computer Engineering
University of Houston
Houston, USA
xli83@central.uh.edu

*Abstract*—Accurate wind speed prediction is crucial for designing and selecting sites for offshore wind farms. This paper investigates the effectiveness of various machine learning models in predicting offshore wind power for a site near the Gulf of Mexico by analyzing meteorological data. After collecting and preprocessing meteorological data, nine different input feature combinations were designed to assess their impact on wind power predictions at multiple heights. The results show that using wind speed as the output feature improves prediction accuracy by approximately 10% compared to using wind power as the output. In addition, the improvement of multi-feature input compared with single-feature input is not obvious mainly due to the poor correlation among key features and limited generalization ability of models. These findings underscore the importance of selecting appropriate output features and highlight considerations for using machine learning in wind power forecasting, offering insights that could guide future wind power prediction models and conversion techniques.

*Index Terms*--Fully Connected Neural Network model, feature combination, machine learning, wind power prediction, wind turbine.

## I. INTRODUCTION

With the ever-increasing energy demand and more severe environmental problems, offshore wind power becomes an important part of the power system, which can supplement both the load demand and near-coast electricity consumption [1][2]. In particular, it powers offshore oil and gas platforms and other subsea loads, replacing traditional diesel generators and contributing to decarbonization. Wind speed and wind power forecasting provide support for the construction of offshore wind farms. However, wind speed and wind power forecasting need to be done based on a large amount of meteorological data and historical data [3]. Additionally, the changes in wind speed characteristics are highly uncertain and instantaneous. To tackle these issues, advanced machine learning (ML) models are essential. It is also necessary to ensure the rationality of the features, the integrity of the data and the rationality of the prediction step in ML models, so data preprocessing and feature selection become very important.

In literature, a large number of ML models have been developed and applied to short-term and long-term regression prediction tasks. Commonly used Prediction models include Support Vector Machine (SVM), Long Short-Term Memory (LSTM) and Transformer model and so on [4][5][6][7]. At present, relatively many new prediction models or methods for optimizing prediction results are proposed such as combining Gaussian Mixture Model (GMM) clustering with a Conditional Generative Adversarial Network (CGAN) based on a CNN-GRU structure has shown high precision in predicting short-term offshore wind farm power generation [8], conducting temporal convolution memory network (TCMN) model realizes wind speed forecast which consists of a long short-term memory network (LSTM) and a temporal convolutional network (TCN) [9], considering spatiotemporal feature mining and error correction improves wind power forecasting and some applications of artificial intelligence (AI) approaches in developing a robust wind power forecast system and intelligent energy management [10][11][12][13].

In addition, for sites where wind turbines are being considered for the first time, wind power data must be calculated using a formula based on wind speed. The choice of input feature combinations and prediction targets can vary during model training, such as using wind speed data or wind power data. For example, if the ML model is used for wind speed prediction and the required wind power data is calculated using the formula and the target feature is wind speed. Alternatively, if the input feature is based on wind speed and the formula is used to calculate the required wind power data and then the wind power data is directly predicted, this time the target feature is wind power [14][15].

This paper takes the site near the Gulf of Mexico as the research object and uses various ML models to perform offshore wind power forecasting tasks based only on meteorological characteristics. The research includes collecting and preprocessing meteorological data, designing different input feature combinations and changing output feature to train ML models and finally evaluating the results. The results show that the prediction accuracy is improved by about 10% when wind speed is used as the output target in the entire wind power forecasting task. The lack of significant improvement in multiple input features compared with single input feature may be due to the poor correlation between key features and limited model generalization ability. These findings highlight the importance of output feature selection and provide insights for future wind power prediction models and conversion technologies.

The rest of this paper is organized as follows. Feature selection and data preprocessing are discussed in Section II. The realization of forecast task based on ML models is detailed in Section III. The method of wind power forecasting is explored in Section IV. The results are analyzed in Section V and Section VI concludes the paper.

## II. FEATURE SELECTION AND DATA PREPROCESSING

### A. Selection of station site and data

The offshore wind power site selection needs to consider factors such as distance from shore, water depth, wind speed and wind stability. This paper selects a site near the Gulf of Mexico. Based on the multi-year average offshore wind speed mapped at 100 meters above sea level as seen in Fig. 1, according to NREL, the average offshore wind speed in the western part of the Gulf of Mexico is relatively high and the water depth is suitable.

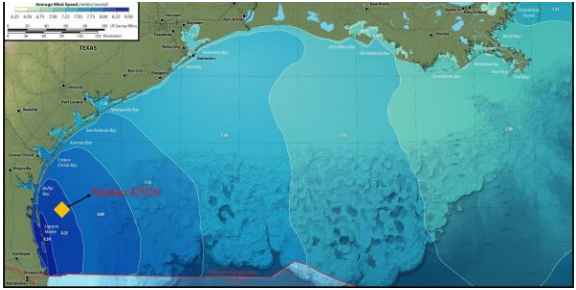

Fig. 1. Offshore wind speed multiyear average at 100 meters above surface level from NREL.

Station 42020 is selected as the research object based on the site information provided by the National Data Buoy Center. The site buoy is a 3 meters discus buoy located 60 nautical miles southeast of Corpus Christi. The site elevation is sea level and the anemometer height is 3.8 meters above the site elevation and the water depth is 86 meters. Standard meteorological data of station from 7/25/2021 22:30 to 10/11/2023 6:00 with 10 minutes interval includes basic time information, wind direction (WDIR), wind speed (WSPD), gust speed (GST), pressure (PRES), air temperature (ATMP), water temperature (WTMP) and dew point temperature (DEWP). The entire dataset consists of 116,071 rows and 12 columns. The repaired data includes 116,254 rows and 12 columns and the time series are complete and continuous.

### B. Analysis of feature correlation coefficient and importance to prediction results

The correlation matrix between WSPD and other features is calculated to perform correlation analysis and feature selection is shown in Fig 2, which aims to reduce the dimension of the data, improve the efficiency and performance of the model and also prevent overfitting problems.

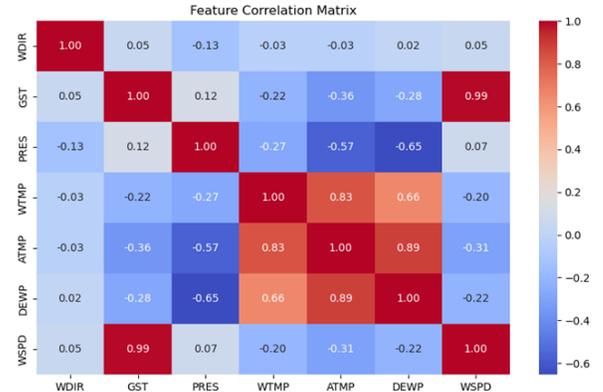

Fig. 2. Diagram of feature correlation matrix.

The correlation between GST and WSPD is as high as 0.99, which indicates that there is a very strong linear relationship between them, which means that GST can be used as an input feature that clearly reflects the behavioral trend of WSPD. Besides, the feature correlation between WDIR and WSPD is low and can be considered for deletion.

There is a strong correlation between ATMP, WTMP and DEWP rather than the target feature WSPD in feature correlation matrix results, which may cause internal redundancy and overfitting problems in the dataset. Specific solutions include deleting some redundant features or merging them into new feature items. Here, deleting features is given priority because the merged new features require deep model training to find the relationship between the new features and the target features, which undoubtedly increases the workload and prolongs the training time.

Based on the Random Forest Regression model, the important analysis of PRES, ATMP, WTMP and DEWP features will be conducted to determine their differences in contribution to the prediction results and the formula is shown as follows.

$$Feature\ Importance(j) = \sum_{t=1}^{T} \sum_{n \in N_t} \frac{I_n(j)}{Total\ Splits} \quad (1)$$

where $T$ is the number of trees in the forest, $N$ is the set of all split nodes in tree $t$ and $I$ represent the impurity reduction when feature $j$ is selected for splitting at node $n$ and Total Splits is the total impurity reduction of split nodes in all trees used for normalization. The final feature importance is the ratio of the sum of the impurity reduction of the feature in all trees to the total impurity reduction [16].

The result shows that DEWP has a lowest feature importance among the temperature features and can be considered for deletion too. The final input features include WSPD, GST, PRES, ATMP and WTMP and output feature is WSPD. The ratio of training set, validation set and test set is 8:1:1 respectively. The training set is used to train the model based on historical data, the validation set is used for model tuning and monitoring and the test set is used to evaluate the model performance to reflect the generalization ability of the model.

## C. Selection of sliding window size

Different prediction modes can be achieved by defining the sliding window size by setting different time steps. The past time step is set to 1, 3, 6, 12 and 24 hours and the forecast time step is set to 10 minutes, 30 minutes, and 1, 3, and 6 hours respectively. Here we assume that the past time step is greater than or equal to the forecast time step. The Fully Connected Neural Network (FCNN) model is used for cyclic training based on the validation dataset to find a better forecast mode that meets practical applications. The reason for using FCNN is that it shortens the training time after flattening the data and as the most basic neural network structure, FCNN is neither complex nor too simple.

TABLE I. COMPARISON OF FCNN MODEL PREDICTION RESULTS UNDER DIFFERENT SLIDING WINDOW SIZE SETTINGS

| Past time | Forecast time | MAE (m/s) | RMSE (%) | MAPE (%) | $R^2$ (%) |
|---|---|---|---|---|---|
| 1h | 10min | 0.2974 | 0.4438 | 11.3736 | 96.13 |
| **3h** | **10min** | **0.3234** | **0.4668** | **11.8816** | **95.71** |
| 6h | 10min | 0.5373 | 0.6725 | 18.1735 | 91.11 |
| 12h | 10min | 0.6382 | 0.7941 | 22.0086 | 87.60 |
| 24h | 10min | 0.4548 | 0.6173 | 16.1489 | 92.51 |

The results show that as the past time step increases, the prediction results for longer time steps will be better. However, if the past time step is close to the prediction time step, the generalization ability of the model will be lost and the results will be poor. Considering the accuracy and practical application of the model prediction results, we use the data of the past 3 hours to predict the wind speed in the next 10 minutes. A subset of selected results, with a look-ahead prediction time of 10 minutes, is presented in Table I. Although the prediction step is relatively short, ensuring the accuracy of the predicted wind speed is beneficial to the subsequent wind power calculation.

## III. REALIZATION OF FORECAST TASK BASED ON ML MODELS

### A. Evaluation indicators

The quality of the wind speed prediction model is measured by evaluation indicators. Mean of Absolute Error (MAE) in test data, Root Mean Squared Error (RMSE), Mean Absolute Percentage Error (MAPE) and the Coefficient of Determination ($R^2$). Besides, Symmetric Mean Absolute Percentage Error (SMAPE) is used to minimize the impact of zero values on the accuracy of the results. The calculation formula of this indicator is shown as follows.

$$SMAPE = \frac{1}{n}\sum_{i=1}^{n}\frac{|y_i - \hat{y}_i|}{\frac{|y_i| + |\hat{y}_i|}{2}} \times 100 \qquad (2)$$

where $y_i$ is the true value, $\hat{y}_i$ is the predicted value, and $n$ is the number of data. All indicator results are obtained based on the predicted results and true values of the test set.

### B. Realization of forecast task

Based on model SVM, GRU, LSTM, FCNN, LCRN and Transformer [17], the offshore wind speed prediction task is realized and the model accuracy test metrics is output for comparison as Table II.

TABLE II. THE RESULTS OF PREDICTION TASK BASED ON VARIOUS MACHINE LEARNING MODELS

| Model | MAE (M/S) | RMSE (%) | MAPE (%) | $R^2$ (%) |
|---|---|---|---|---|
| SVM | 0.2849 | 0.4261 | 8.574 | 96.43 |
| GRU | 0.2859 | 0.4268 | 9.567 | 96.42 |
| LSTM | 0.2879 | 0.4235 | 9.237 | 96.47 |
| FCNN | 0.3234 | 0.4668 | 11.88 | 95.71 |
| LCRN | 0.3462 | 0.4876 | 15.33 | 95.32 |
| Transformer | 0.3769 | 0.5925 | 13.20 | 93.09 |

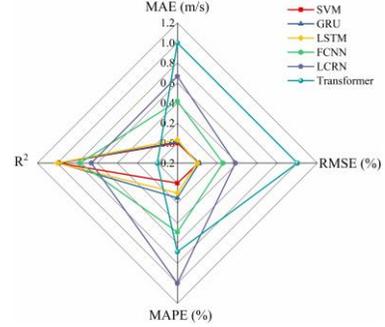

Fig. 3. Radar chart of prediction task results.

The results show that complex neural networks do not produce good results for wind speed prediction. This may be because that the 10 minutes prediction time is a short-term prediction. The SVM and GRU models perform better for short-term predictions. From the radar chart as shown in Fig.3, SVM, GRU and LSTM perform better and the higher determination coefficient of LSTM proves that the model has stronger generalization ability. However, the performance of FCNN, LCRN and Transformer models is poor.

## IV. WIND POWER FORECASTING METHODS

### A. Calculation and conversion of wind power

Based on the logarithmic wind profile formula, the wind speed at 100 meters can be calculated from the wind speed at 3.8 meters. The formula can be expressed as follows.

$$v_2 = v_1 \frac{ln\left(\frac{h_2}{z_0}\right)}{ln\left(\frac{h_1}{z_0}\right)} \qquad (3)$$

where $h_2$ is the height of wind turbine blade which is equal to 100 meters above sea level and $h_1$ is the height of anemometer which is equal to 3.8 meters above sea level. $v_2$ is the average wind speed at 100 meters above sea level and $v_1$ is the average wind speed at 3.8 meters above sea level and $z_0$ is the roughness length of the sea which is equal to 0.0002 meters [18].

Wind energy is the total energy that can be captured by the wind turbine blades, while wind power is the total energy rate that the wind turbine extracts from the wind. If wind power is converted into electrical power, the actual efficiency and losses of the wind turbine and power generation system must also be considered, which will not be discussed in depth here. The average wind speed at 100 meters above sea level are shown as follows.

$$P = \frac{1}{2} \times \rho \times A \times v_2^3 \times C_p \qquad (4)$$

where $\rho$ is the air density at the tower height, $A$ is the swept area of the wind blade, $v_2$ is the wind speed at the tower height and $C_p$ is the efficiency of electricity generation. Here the air density at 100 meters above sea level is about 1.2kg/m$^3$ and set the assumption of wind turbines. Assuming the rotor diameter is 150 meters, the cut-in wind speed is 3m/s, the cut-out wind speed is 25m/s, the rated wind speed is 12-14m/s and the rated power is 8MW, the use of floating foundation to construct the wind turbine is also considered.

### B. Wind turbine wind speed and power range

Based on the cut-in, cut-out wind speed and the rated wind speed, the operating range can be set of the wind speed at 3.8 meters above sea level and 100 meters above sea level respectively. The start wind speed range in equation (5)-(7), rated wind speed range in equation (8)-(10) and cut-off wind speed range in equation (11)-(13) are shown as follows.

$$2.3 m/s \leq WSPD\ at\ 3.8m \leq 9.3 m/s \quad (5)$$
$$3 m/s \leq WSPD\ at\ 100m \leq 12.4 m/s \quad (6)$$
$$0 MW \leq Wind\ Power \leq 8 MW \quad (7)$$
$$9.3 m/s \leq WSPD\ at\ 3.8m \leq 18.8 m/s \quad (8)$$
$$12.4 m/s \leq WSPD\ at\ 100m \leq 25 m/s \quad (9)$$
$$Wind\ Power = 8 MW \quad (10)$$
$$WSPD\ at\ 3.8m \geq 18.8 m/s \quad (11)$$
$$WSPD\ at\ 100m \geq 25 m/s \quad (12)$$
$$Wind\ Power = 0 MW \quad (13)$$

Based on wind speed data analysis, approximately 92.8% of wind energy can be converted into electricity. This proves that the wind speed utilization rate of this site is relatively objective.

### C. Model training with different input feature combinations

In Fig. 4, the input feature combinations of the training model are classified into nine cases. The differences between the three modes are in the number and type of input features and the output features, and the difference between the nine cases lies in whether the key feature is wind speed or wind power. The green box in Fig. 4 is the input part and the red box is the output part.

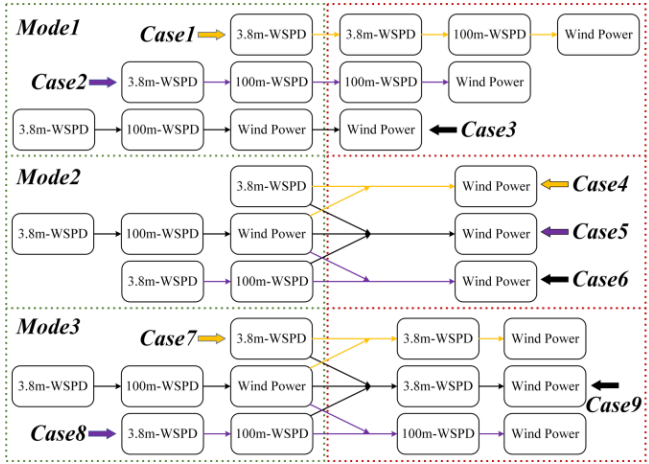

Fig. 4. Diagram of input feature combination pattern under nine cases.

## V. RESULTS ANALYSIS

Based on six ML models, nine different input feature combination cases are used for prediction tasks and the results are compared. The model performance indicators MAE, RMSE, SMAPE and Coefficient of Determination shows that the accuracy remains nearly the same in pairs of cases such as Case1 vs. Case2, Case4 vs. Case5, and Case7 vs. Case8, because the behavior of wind speeds at 3.8m and 100m is similar, differing only in scale. However, the LCRN model has the ability to capture long-term dependencies and spatial nuances and be more sensitive to minor differences which potentially leads to slight variations in predictions and occasional overfitting.

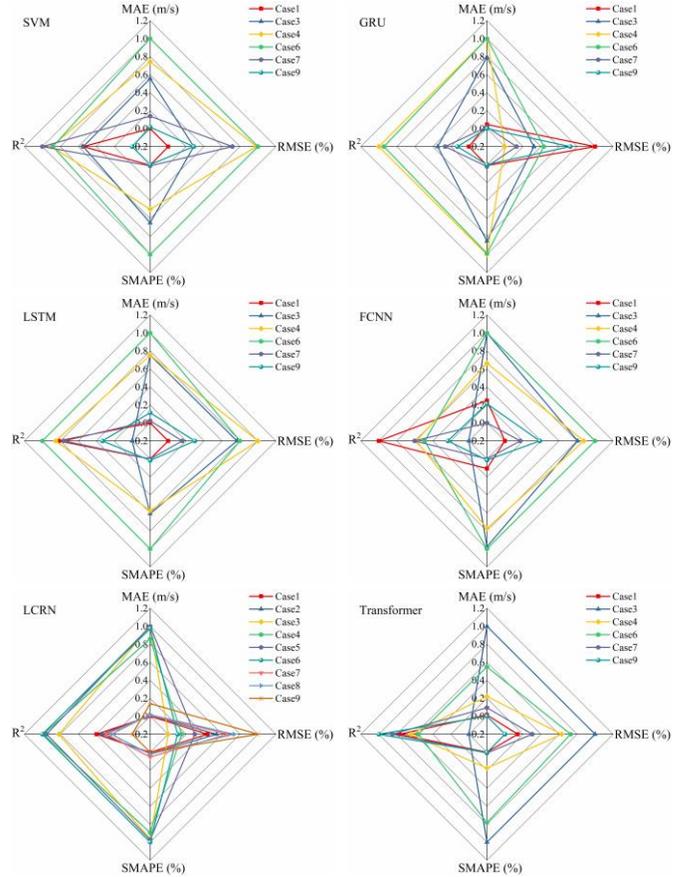

Fig. 5. Radar chart of prediction results for 9 different input feature combination cases.

In Mode 1, results are similar whether using 3.8m or 100m wind speed as input where evaluation compares calculated and actual power values. However, when inputting wind power only, accuracy decreases as the model lacks wind speed information which contributes more effectively to wind power prediction. In Mode 2, evaluation based on predicted versus actual power shows that including either 3.8m or 100m wind speeds alongside power gives similar accuracy. However, adding both wind speeds with wind power in Case 6 reduces accuracy due to data redundancy impacting the model's generalization. In Mode 3, using multiple features to predict wind speed yields better accuracy than predicting wind power. Additionally, predicting wind speed with multiple inputs performs comparably to cases where both input and output are wind speeds, reinforcing the advantage of direct wind speed modeling over indirect wind power predictions. Based on the results, the radar chart is drawn to compare the results of the same model under different input feature combinations. In order to more intuitively compare the performance indicators under different cases, the results of each indicator are normalized and duplicate results are deleted.

Fig. 5 shows that when the output feature is wind speed, the prediction accuracy is better than when the output feature is wind power. In addition, simpler models such as SVM and GRU have higher prediction accuracy and multi-feature input does not improve the prediction results. Combining the above two points, it can be concluded that considering the complexity of the data set and the relevance of the input features to the target, the accuracy of the wind power prediction results is closely related to the accuracy of the wind speed prediction results.

## VI. Conclusion

This paper investigates the prediction of offshore wind power. The offshore site is located near the Gulf of Mexico. Site 42020 is selected as the research site based on the average wind speed and the integrity of meteorological data over the years. Since each meteorological feature contributes differently to the prediction results, feature selection and data preprocessing are performed. Six machine learning models with different complexities are selected to implement the regression prediction task. Nine different input feature combinations are designed according to the wind speed conversion at different heights and wind power calculation, and then the results show that predicting wind speed as the output can improve accuracy by about 10% over predicting wind power directly, indicating that additional wind power-specific data or more detailed wind speed inputs may be needed for better wind power predictions. The benefit of multiple feature inputs is minimal, probably due to data redundancy. This distracts the model from focusing on key features and a more complex multi-objective model may help. In practice, accurate and timely data collection remains a challenge. Although the model is effective in offline testing, it needs to be frequently updated to ensure adaptability and accuracy in real-time applications.

## Acknowledgment

This project was paid for [in part] with federal funding from the Department of the Treasury through the State of Texas under the Resources and Ecosystems Sustainability, Tourist Opportunities, and Revived Economies of the Gulf Coast States Act of 2012 (RESTORE Act). The statements, findings, conclusions, and recommendations are those of the author(s) and do not necessarily reflect the views of the State of Texas or the Department of the Treasury. Research reported in this paper was also in part supported by an Early-Career Research Fellowship from the Gulf Research Program of the National Academies of Sciences, Engineering, and Medicine. The content is solely the responsibility of the authors and does not necessarily represent the official views of the Gulf Research Program of the National Academies of Sciences, Engineering, and Medicine.